\documentclass[aps,reprint,superscriptaddress,nofootinbib,showpacs]{revtex4-1}
\usepackage{amsmath,latexsym,amssymb,hyperref,graphicx}
\usepackage{slashed}

\hypersetup{colorlinks,citecolor= nicegreen,linkcolor= nicered}
\usepackage{color}
\definecolor{nicered}{rgb}{0.7,0.1,0.1}
\definecolor{nicegreen}{rgb}{0.1,0.5,0.1}

\newcommand\GeV{\text{GeV}}
\newcommand\TeV{\text{TeV}}

\newcommand{\beq}{\begin{equation}}
\newcommand{\eeq}{\end{equation}}
\newcommand{\bea}{\begin{eqnarray}}
\newcommand{\eea}{\end{eqnarray}}

\newcommand\SEC[1]{\medskip\noindent{\sl\bfseries #1}}
\renewcommand{\figurename}{\renewcommand\baselinestretch{1.15}\scriptsize\em Fig.{}}
\renewcommand{\tablename}{\renewcommand\baselinestretch{1.15}\scriptsize\em  Table}

\begin{document}
%%%%%%%%%%%%%%%%%%%%%%%
\addtolength{\belowdisplayskip}{-.5ex}       %   comando per recuperare spazio verticale dopo le formule
%%%%%%%%%%%%%%%%%%%%%%%
\addtolength{\abovedisplayskip}{-.5ex}       %   comando per recuperare spazio verticale prima delle formule
%%%%%%%%%%%%%%%%%%%%%%%

\title{Three Extra  Mirror or Sequential Families: \\a Case for  Heavy Higgs and Inert Doublet }
 
\author{Homero Mart\'inez}
\affiliation{CEA, Saclay, DSM-IRFU-SPP, France}

\author{Alejandra Melfo}
\affiliation{ICTP, Trieste, Italy}
\affiliation{U. de Los Andes, M\'erida, Venezuela}

\author{Fabrizio Nesti}
\affiliation{U. di Ferrara, Ferrara, Italy}

\author{Goran Senjanovi\'{c}}
\affiliation{ICTP, Trieste, Italy}
 
\date{\today}

\begin{abstract}
  \noindent  
  We study the possibility of the existence of extra fermion families
  and an extra Higgs doublet.  We find that requiring the extra Higgs
  doublet to be inert leaves   space for three extra families, allowing for mirror fermion families
  and a dark matter candidate at the same time. The emerging scenario
 is very predictive: it consists of a Standard Model Higgs boson,
  with mass above 400\,\GeV, heavy new quarks between 340 and
 500\,\GeV, light extra neutral leptons, and an inert scalar with a
  mass below $M_Z$.
  \end{abstract}

%12.60.Cn 	Extensions of electroweak gauge sector
%12.60.Fr 	Extensions of electroweak Higgs sector 
%14.65.Jk 	Other quarks (e.g., 4th generations) 
%14.60.Hi 	Other charged heavy leptons 
%14.60.St 	Non-standard-model neutrinos, right-handed neutrinos, etc. 
%14.80.Ec 	Other neutral Higgs bosons 
%14.80.Fd 	Other charged Higgs bosons 


\maketitle

%\noindent   
\looseness=-1% questo per risparmiare una linea che era brutta.
\SEC{Introduction.} It may not be well known that the idea of parity restoration in weak interactions is as old as the suggestion of its breakdown. In their classic paper, Lee and Yang~\cite{Lee:1956qn} proposed the existence of what we will call mirror fermions, so as to make the world left-right symmetric at high energies. By this they meant another version of the proton and the neutron, with opposite chirality under weak interactions.\footnote{Since then, a number of different definitions of mirror particles have been used in the literature. For a recent review, see~\cite{Okun:2006eb}.} Besides the wish to make the world symmetric, there are a number of important theoretical frameworks that imply them: Kaluza-Klein theories~\cite{Witten:1981me}, family unification based on large orthogonal groups~\cite{GellMann:1980vs,Senjanovic:1984rw,Bagger:1984rk}, N=2 supersymmetry~\cite{delAguila:1984qs}, some unified models of gravity~\cite{GGUT}. Mirror fermions appear naturally in the simplest and most physical way of gauging baryon and lepton number symmetry.  Moreover, since they necessarily obtain their masses through the same electroweak symmetry breaking providing ordinary fermion masses, standard perturbativity arguments require their mass scale to lie below 600\,GeV or so~\cite{Marciano:1989ns}.  Since this energy range will be probed soon at the Large Hadron Collider (LHC), an updated investigation on their possible existence is called for.

On the other hand, one of the most appealing features of the Standard Model is the chiral nature of quarks and leptons, and the cancellation of anomalies through their precise matching. With mirror fermions this is gone, as is the understanding of the smallness of fermion masses. Namely, the gauge-invariant mass terms between ordinary and mirror fermions, unless suppressed, would make them pair off and disappear from the low energy world. This implies an approximate symmetry that forbids these terms, often called {\it mirror parity}. In spite of these phenomenological drawbacks, mirror fermions remain a fascinating possibility. This letter is devoted to asses their compatibility with recent experimental constraints and high precision tests of the Standard Model. Notice that with regard to high precision analysis they behave exactly as ordinary fermions, and thus a reader who is uncomfortable with the above setbacks can view our study as referring to the more general question of whether the SM can host three (or more) extra families.

If one defines the SM by its structure, i.e.\ by the quantum numbers of particles in its minimal version, the two central issues that it faces in view of the upcoming LHC probe are the number of families and the number of Higgs doublets. It is interesting that high precision tests constraints can provide a link among them, as it has been noted in~\cite{He:2001tp}. Indeed, as we shall see below, extra families can be reconciled with the most recent bounds from colliders and with precision tests by invoking an extra Higgs doublet. This, we emphasize, implies an addition to the SM that still preserves its fundamental structure.

\looseness=-1%
The existence of the fourth chiral family of quarks and leptons is an old question~\cite{Frampton:1999xi}.  More than ten years ago, it was argued~\cite{Maltoni:1999ta} that it was in accord with the high precision study, and soon after, it was pointed out that it went hand in hand with the heavier Higgs~\cite{He:2001tp}.  Unfortunately, it kept being claimed unacceptable by PDG for years. The case was reopened in recent years by~\cite{Kribs:2007nz}, who argued again that it was perfectly possible, and that it fitted nicely with a heavier Higgs. Since then it has become the subject of intense study~\cite{Holdom:2009rf}. However, it is still believed by and large that high precision tests leave no room for more families beyond the fourth. This is true only if there is no new physics whatsoever in the TeV energies. Already in~\cite{He:2001tp}, it was pointed out that a second Higgs doublet would suffice to accommodate even three extra families of quarks and leptons, be they mirrors or not. It was even claimed~\cite{Novikov:2001md} that the same could be achieved without an extra doublet, at the 2$\sigma$ level, but by allowing quark masses below 200\,\GeV\ (not acceptable any more).
 
In view of the new severe bounds on quark masses from direct searches at Tevatron, these studies must be revisited. Moreover, Tevatron~\cite{Aaltonen:2010sv} has recently placed a new lower bound on the Higgs mass with the inclusion of the fourth family, which becomes even stronger as the number of extra families grows. The point is that the main production process for the Higgs, the gluon fusion, gets amply increased by the addition of each family. The whole set-up becomes rather constrained and it is far from obvious that extra families are still allowed at all.
 
Thus the result that extra three (and no more) families are in agreement with high precision data (with the inclusion of an extra scalar doublet), we find both fascinating and surprising. Moreover, the SM Higgs scalar must weigh more than 400\,GeV or so, in full accord with the increased limits, and in an ideal range for the LHC search. The second doublet, on the other hand, tends to be inert~\cite{Deshpande:1977rw} and light, thus offering a hope for a dark matter candidate~\cite{Barbieri:2006dq, DM}. These important results have a great degree of urgency since they are likely to be tested at LHC already by the end of this year.

\medskip

\SEC{Fermions.} Charged leptons must satisfy the bound $m_E \gtrsim 102.6\,\GeV$ if they are long-lived at LEP, and slightly less if they are short lived~\cite{Amsler:2008zzb} (possible signatures have been recently addressed in~\cite{Carpenter:2010bs}). Neutral leptons on the other hand can be at a much smaller scale if they are stable, a fact used already in~\cite{Maltoni:1999ta} (and revived recently in~\cite{Murayama:2010xb}) to enlarge the parameter space for extra families allowed by high precision tests. If the extra neutral leptons are long lived, their masses are just bounded by the invisible $Z$ width to be $m_N \gtrsim 45\,\GeV$~\cite{Carpenter:2010sm} (limits can be more stringent for large mixing angles with ordinary leptons~\cite{Carpenter:2010dt}, not assumed here). The $S$-$T$ contributions from leptons are found to be minimal for smaller neutral lepton masses, and for a mass ratio of charged to neutral leptons in the range 1.5--3. We start from the range
\beq
m_E: [100 - 300]\,\GeV \qquad m_N : [50 - 250]\,\GeV\,.
\label{leptons}
\eeq
Tevatron lower limits on fourth generation quark masses are constantly improving, but are obviously dependent on unknown mixings with the ordinary quarks. To be on the safe side, we have adopted the most stringent lower limit from CDF~\cite{Aaltonen:2009nr} on direct searches for down type quarks, namely $m_D \gtrsim 338\,\GeV$, noting that this can be lowered down to $249\,\GeV$ for the long-lived case~\cite{Aaltonen:2009kea}.  For large quark masses above $350\,\GeV$, the parameters $S$, $T$ do not strongly constrain the quark masses but rather require $m_U/m_D\sim 1.1$, in order to avoid a too large $T$ parameter. We have therefore taken the extra quark doublet components to be almost degenerate, with $m_U \gtrsim m_D $. With large quarks masses, Yukawa couplings become non-perturbative very fast. Setting a cutoff for the scale of new physics at 1\,\TeV, extra quarks have to be lighter than about 450\,\GeV. These limits will be refined below, when we have more information about the Higgs masses. We start with the range
\beq
  \label{quarks}
   m_D, m_U: [340 - 500]\,\GeV  \qquad m_D \sim m_U\,. 
\eeq

\SEC{Scalars.} We denote by $(C,A,S,h)$ the charged, neutral CP odd, and two neutral CP even states respectively. Their contribution to the oblique parameters depends also on the ratio of vevs $\tan \beta = v_2/v_1$ and on  the angle of rotation into the CP even neutral eigenstates, $\alpha$. These quantities appear only in the combination $\beta-\alpha$, and one can see immediately from the analytical expressions  in~\cite{He:2001tp} that  $\chi^2$ calculated from $S$ and $T$ is extremized for the values $\beta-\alpha = 0, \pi/2$. 
Exploring the range  $100 - 600\,\GeV $  of the doublet component masses   for these two values, the minimum is found to be at $\beta = \alpha $. In other words, the situation where $h$ is the SM Higgs and $C,A,S$ form an inert doublet is preferred by the high precision constraints.

Having established that models with extra families prefer the second Higgs doublet to be inert, we can now identify several restrictions imposed on the mass scales of the scalars involved.  In principle the invisible $Z$ width only restricts $m_S + m_A \gtrsim M_Z$. However, in~\cite{Lundstrom:2008ai}, searches for neutralinos in LEP II are translated  to inert doublet components. For a light $S$ with mass $m_S \lesssim M_Z $, LEP II excludes $m_A \lesssim 100 -  120\,\GeV$. We shall see that a low mass $S$ is preferred by high precision also. Chargino searches can also be translated to give a limit on the charged scalar, $m_C \gtrsim 70\,\GeV$~\cite{Pierce:2007ut}. Finally, a lower limit on the mass of $S$ will come from the four-body decay  $Z \to  S \, S \, Z^* \to S \, S \, f\,  f $,  to be on the safe side we set $m_S \gtrsim 50\,\GeV$.  

The most recent analysis by Tevatron~\cite{Aaltonen:2010sv} has excluded a SM Higgs with a mass between $131\,\GeV$ and $200\,\GeV$ in the presence of a fourth family, from the upper limits on gluon-fusion production (enhanced by a factor of 9) and decay into $W$: $g \, g \to h \to W^+ \, W^-$.  Three extra families enhance the Higgs production by a factor of 49, and one can easily check (see for example~\cite{Djouadi:2005gi}) that $h \to W^+ \, W^-$ is still largely the dominant decay for a Higgs heavier than about $200\,\GeV$. The corresponding Tevatron exclusion on the Higgs mass can be estimated roughly as $M_h>300\,\GeV$. 

However with extra quarks, stability becomes a concern. In the SM, the one-loop RGE for the quartic Higgs coupling $\lambda$ reads (see e.g.~\cite{Djouadi:2005gi})
\bea
\frac{d\lambda}{d \log Q^2} &\simeq & 
\frac{1}{16 \pi^2} \left[ 12 \lambda^2 + 6 \lambda y_t^2 - 3 y_t^4 -   \frac{3}{2} \lambda (3 g_2^2 + g_1^2) \right. \nonumber  \\ 
& & \left . \qquad \quad{}+ \frac{3}{16} \big( 2 g_2^4 + (g_2^2 + g_1^2)^2 \big)\right],\ \ 
\label{rge}
\eea
where $g_1$, $g_2$ are the $U(1)$ and $SU(2)$ gauge couplings and $y_t$ is the top Yukawa.  With new families the dominant contribution, of the same form, comes from both up and down heavy quarks.  Also, there are additional couplings among the two scalar doublets. Their contribution is subleading~\cite{Barbieri:2006dq} if the SM Higgs is heavy.  The analysis in this case shows that with three extra families with quark masses $\sim 340\,\GeV$, the SM Higgs must be above $\sim 400 (350) \,\GeV$, if the cutoff scale is set at $1 (0.7)\,\TeV$.  On the other hand, a Higgs boson in the light mass window $ 115 < M_h < 130 $ can be much more problematic for stability, particularly if the doublet masses are very split so that the contribution of the additional couplings to~\eqref{rge} cannot be neglected. We shall see below that this is precisely the case.

Perturbativity certainly sets an upper limit on Higgs masses, but it is not straightforward to translate it into a definite scale.  In~\cite{Barbieri:2006dq}, a bound of 600\,\GeV\ is chosen also for the extra doublet masses, whose running is milder.  We find a similar limit for the SM Higgs with three families. (In this we differ from~\cite{Anber:2009tz}, where scalar masses above the \TeV\ scale are allowed.)  

\begin{figure}
\centerline{\includegraphics[width=.85\columnwidth]{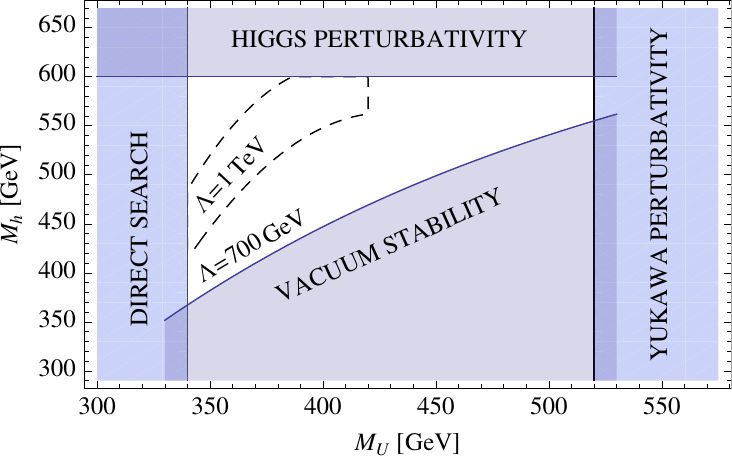}}%
\vspace{-2.5ex}%
\caption{Stability and perturbativity limits on $m_h$, $m_U$ (shaded regions) in the presence of three extra generations of heavy quarks (as in table~\ref{table}, {\bfseries\slshape (i)}. The allowed region (white) corresponds to a low cut-off $\Lambda=700\,\GeV$; the dashed-contour region to $\Lambda=1\,TeV$.\label{stability}}
\vspace{-2ex}%
\end{figure}
In Figure~\ref{stability}, we summarize the constraints from perturbativity and stability for a heavy SM Higgs and heavy quark masses, resulting from the coupled RG running, for three extra families.  These limits become much more restrictive for additional extra families.  Notice incidentally that when requiring perturbativity up to $\Lambda=\TeV$, for higher fermion masses the upper bound on $m_h$ is alleviated by the larger Yukawa couplings. We consider thus
\bea
m_C &:& [100 - 600]\,\GeV , \;  m_h:[115-130]  ,  [400 - 600]\,\GeV \,.\nonumber \\
m_S &:& [ 50 - 600 ]\,\GeV ,  \; \; m_A: [100 - 600]\,\GeV ,  \ \ \  \beta = \alpha  
\label{scalars}
\eea

\SEC{Allowed space.}  We have performed a scan of this parameter space, first by varying the masses in the ranges (\ref{leptons}), (\ref{quarks}), (\ref{scalars}) in intervals of $30\,\GeV$, with families differing in mass by $10\,\GeV$, calculated their contributions to the parameters $S$, $T$ and $U$, and selected them according to the experimentally allowed regions from Ref.~\cite{gfitter}.  We then refined the search by generating random points in the most promising regions in $S$-$T$-$U$ space.

We find that the inclusion of a third family with heavy quarks puts some strain in the high precision variables, and in particular on the extra neutral leptons masses: there are no allowed points within the 95\% C.L. regions in $S$-$T$-$U$ space for neutral lepton masses above $ \sim 200\,\GeV$. The extra quark masses do not have an impact for this high mass range, as long as the doublets are almost degenerate.

Two different possibilities emerge from the analysis: {\bf (i)}: $m_h \gtrsim 400 GeV$, and {\bf (ii)}: $115 GeV \lesssim m_h \lesssim 130 GeV$. Case {\bfseries (ii)}, although it provides the bests fits with high precision data, is extremely difficult to reconcile with vacuum stability requirements to a reasonably high cutoff, the only possibility being that some of the inert doublet scalars have very large masses, with unacceptably large couplings.
%This issue requires a more careful and complete analysis of the RGE equations, but one can expect this case to be inviable.

\begin{figure}%
\vspace*{-1ex}
\centerline{\includegraphics[width=.98\columnwidth]{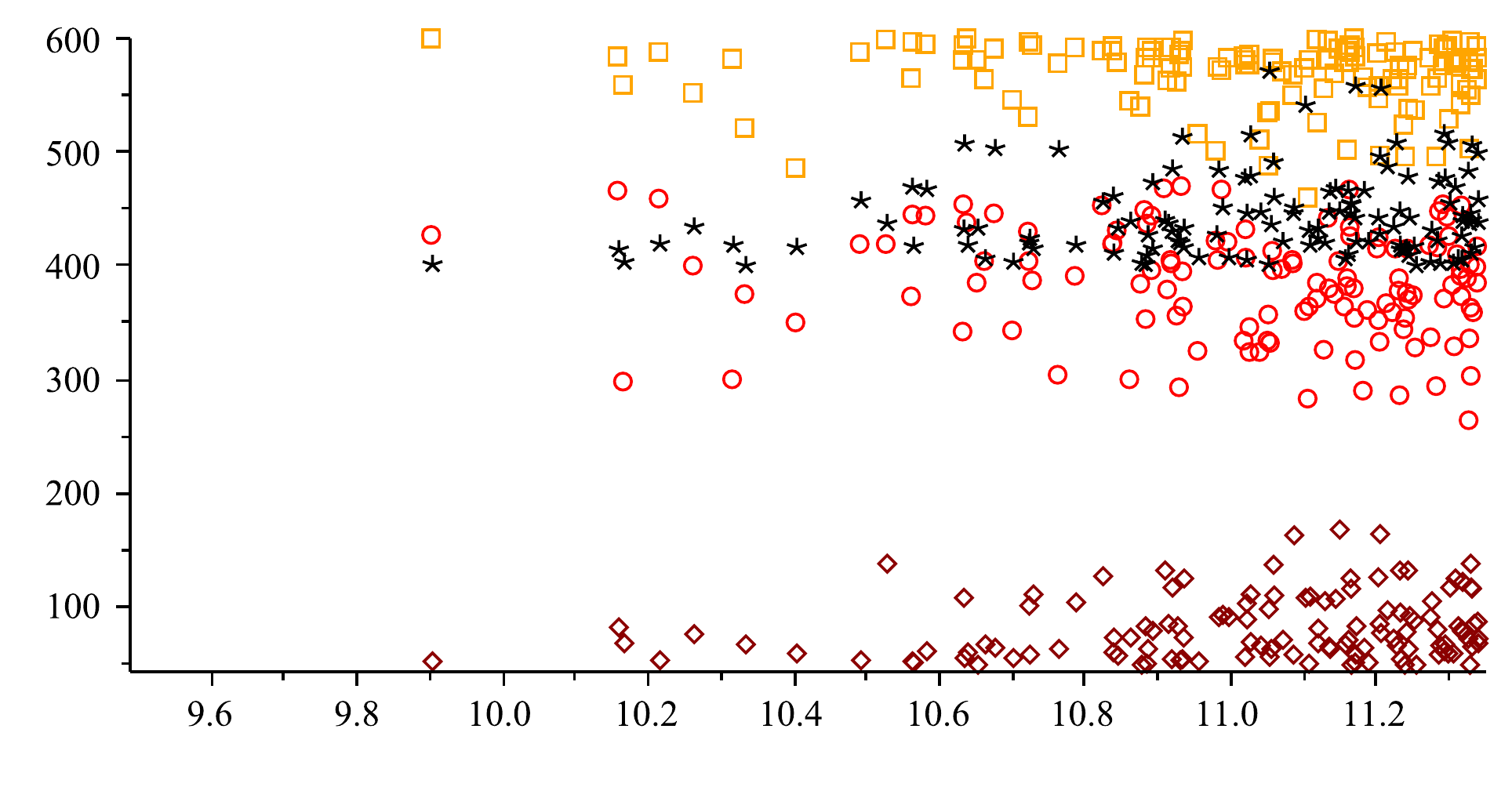}}%
\vspace{-4ex}%
\caption{Allowed mass ranges of the scalars $C$ (red circles), $A$ (orange squares), $S$ (brown diamonds) and  $h$ (black asterisks)  within the 99\% C.L. region in  $S$-$T$-$U$ space for randomly generated points, with a heavy SM Higgs boson.  Masses in \GeV\ are plotted against $\Delta\chi^2$. \label{fig1}}
\vspace{-4ex}%
\end{figure}

Case {\bfseries (i)} on the other hand, is allowed by perturbativity and stability requirements, as show in Fig..~\ref{stability}.  It is possible to accommodate three extra families below the 99\% CL, with the best fit point at 2.3 $\sigma$.  This implies that adding extra families would put the SM Higgs within the reach of experiments in the very near future. In this case $m_S$ has relatively small values, and the best fits prefer a low mass, $m_S \lesssim M_Z$, in accordance with its proposed role as dark matter.  The neutral scalar $A$ is extremely heavy, while $C$ lies in the intermediate mass range. Due to the similar contributions of $A$ and $S$ to the high precision parameters, these results are preserved under interchange of their roles, {\em i.e.}\ a solution with light $A$ and heavy $S$ also exists.  We report these results in Table~\ref{table} and Fig.~\ref{fig1}.

\begin{table}             
\begin{tabular}{ll}
   \begin{tabular}{|c|c|c|}
   \hline
& case {\bfseries (i)} & case {\bfseries (ii)}  \\
 \hline                                                        
  $m_C$ & 250 -- 500 &   180 -- 530    \\              
   $m_A$ & 450 -- 600 & 380 -- 600   \\                  
   $m_S$ & 50 -- 80 &  50 -- 200    \\                  
   $m_h$ & 400 -- 600 &  $\sim $120    \\             
 %  $ m_N$ & 50 -- 200 &   50 -- 200   \\              
  % $m_E$ &100 --300  &  100 -- 300   \\              
  % $m_{U,D}$\ \  & 340 -- 500 &  340 -- 500  \\
\hline
\end{tabular}     &
\   \begin{tabular}{|c|c|c|}
   \hline
& case {\bfseries (i)} & case {\bfseries (ii)}  \\
 \hline                                                        
%  $m_C$ & 250 -- 500 &   180 -- 530    \\              
%   $m_A$ & 450 -- 600 & 380 -- 600   \\                  
%   $m_S$ & 50 -- 80 &  50 -- 200    \\                  
%   $m_h$ & 400 -- 600 &  $\sim $120    \\             
  $m_N$ & 50 -- 200 &   50 -- 200   \\              
  $m_E$ &100 --300  &  100 -- 300   \\              
  $m_{U}$\ \  & 340 -- 500 &  340 -- 500  \\
   $m_{D}$\ \  & 340 -- 500 &  340 -- 500  \\
\hline
\end{tabular}  \\
  \end{tabular}                                                                                                                
\vspace{-1ex}%
\caption{Allowed scalar and fermion masses in \GeV\ within different  $S$-$T$-$U$ contours at 99\% C.L., scanning parameters in 30\,\GeV\ steps.\label{table}}. 
\end{table}

\SEC{Even more (mirror) families?} Again, their existence depends on the possible presence of other physical states and interactions. Let us still stick to only an additional scalar doublet, now that we saw that it can be a natural candidate for the dark matter, and ask if we may have an additional mirror and ordinary family (or say eight families altogether). The case of four normal and four mirror families was discussed~\cite{Bagger:1984gz} in the $SO(18)$ based unified theory of families. Perturbative unification with a desert, however, requires an upper bound on new fermion masses~\cite{Bagger:1984gz} of $ \sum m_Q^2 \lesssim (350\,GeV)^2 $, and is in clear contradiction with the new Tevatron lower bounds on $m_Q$.  A possible way out could be intermediate mass scales that could prevent the Yukawa couplings from getting strong.

\begin{figure}
\centering{%
\!\!\includegraphics[width=.33\columnwidth]{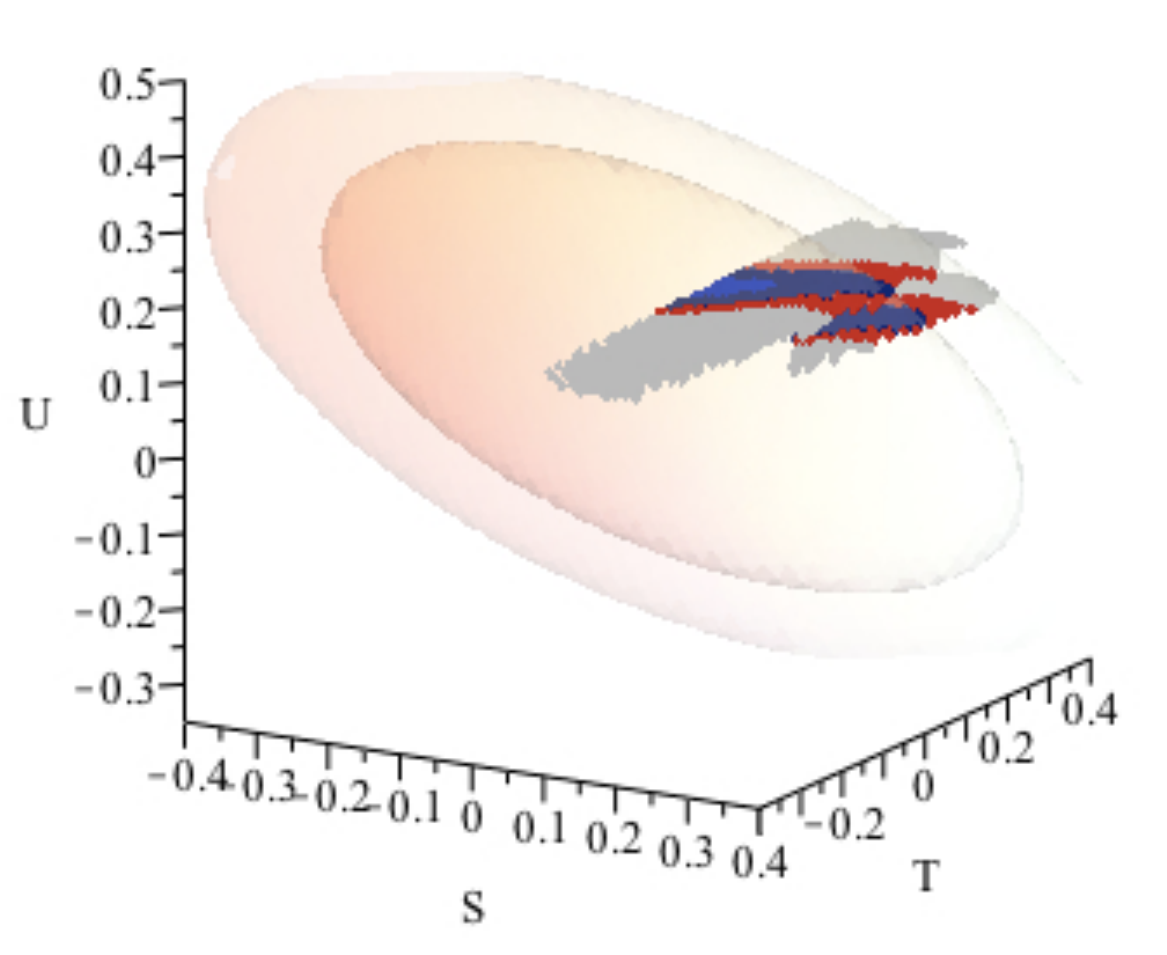}%
\includegraphics[width=.33\columnwidth]{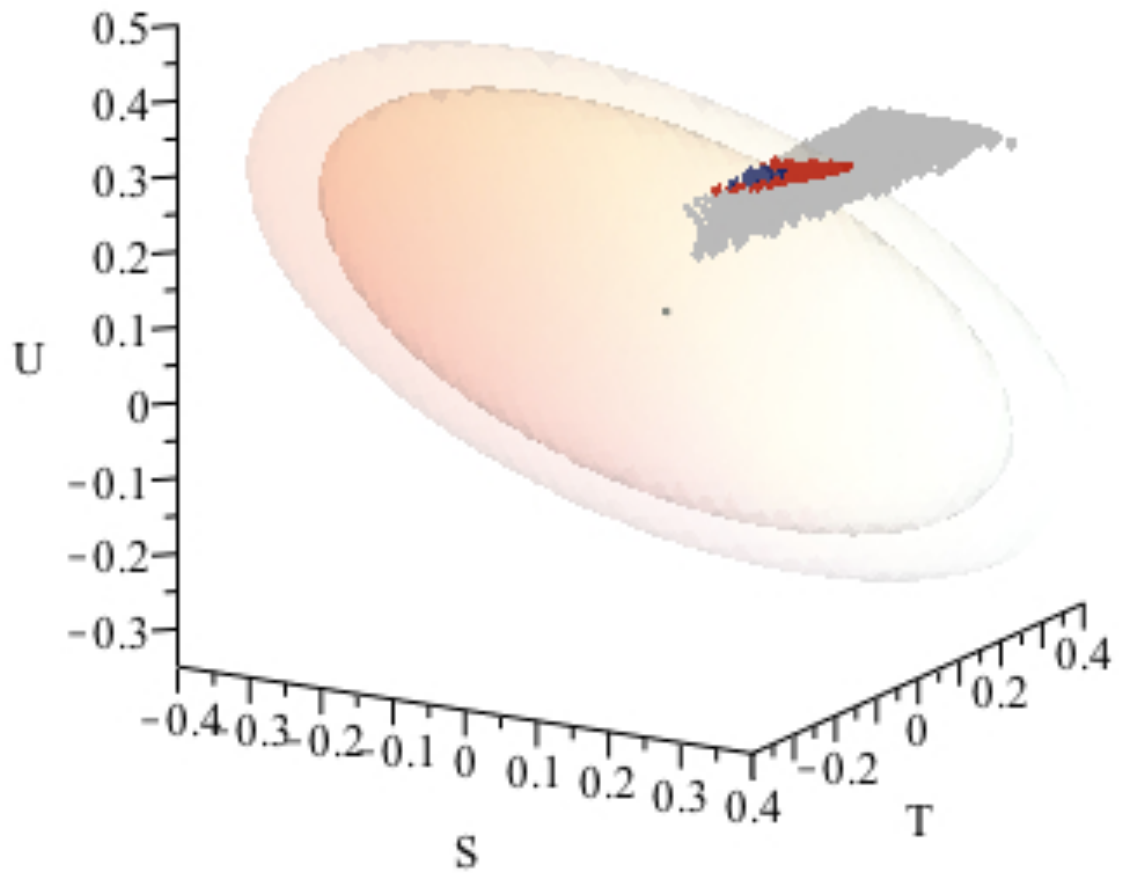}%
\includegraphics[width=.33\columnwidth]{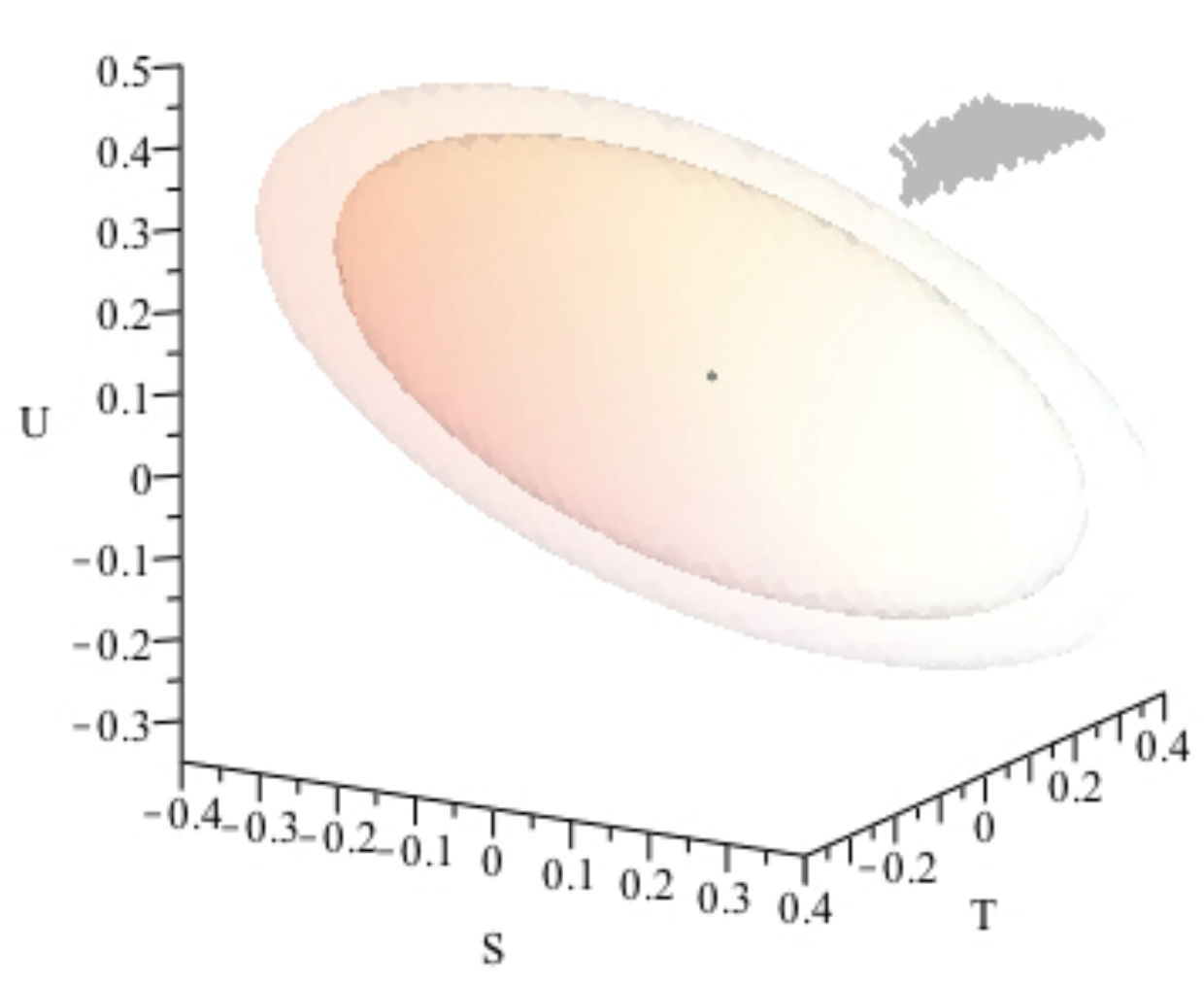}%
}%
\vspace{-2ex}
\caption{Points intercepting the allowed region in $S$-$T$-$U$ space, for 2, 3, 4 extra families. Blue (dark grey) and red (black) points are allowed at  95\%, 99\% CL, and the corresponding contours are shown. The grey points are not allowed.\label{figELL}}
\vspace{-4ex}%
\end{figure}

In order to avoid any theoretical prejudices, we studied the possibilities of having any number of additional families, from two to five (one extra family works nice even with the single Higgs doublet).  We find that the constraints on the parameter $U$ are crucial in this case.  The details of this investigation will be reported elsewhere~\cite{long}; here we will only quote the results.  Not surprisingly, two extra families are allowed for a large range of particle masses, as long as another Higgs doublet is present. Whereas as we have seen three extra families can still exist,  four extra families are excluded at the 99\% C.L. (see figure~\ref{figELL}).  This conclusion favors a particular version of the  $SO(18)$ augmented with a Peccei-Quinn symmetry for it can lead to only three normal and three mirror families at low energies~\cite{Senjanovic:1984rw}.

\SEC{Concluding remarks.} To summarize, we have shown the existence of three (mirror or sequential) families is still perfectly in accord with the SM, as long as an additional Higgs doublet is also present. Moreover, the extra doublet prefers to be inert with  its real neutral component in the light mass range, becoming a candidate for the  dark matter particle. The SM Higgs in this case has to weigh more than 400\,\GeV, still in the perturbative regime. Low scale perturbativity restricts the extra quark masses to the narrow interval $\sim [350-500]\,\GeV$.  At higher scales, new physics is expected to intervene at less than 2\,\TeV\ (for recent studies see~\cite{str}). This exciting scenario is easily testable at LHC.  

\SEC{Acknowledgments.}  We thank Martin Goebel for useful correspondence, and Miha Nemev\v sek and Yue Zhang for discussions. We are deeply grateful to Abdelhak Djouadi for his interest, encouragement, useful comments and a careful reading of the manuscript.  F.N. wishes to thank ICTP for hospitality during the completion of this work.

\end{document}